\documentclass[
reprint,
nofootinbib,
amsmath,amssymb,
aps,
pra,
twocolumn,
floatfix,
]{revtex4-2}

\usepackage{algorithm}
\usepackage{amssymb}   
\usepackage{amsfonts}
\usepackage{amsmath}
\usepackage{mathtools}
\usepackage{MnSymbol}
\usepackage{dsfont}
\usepackage{dcolumn}   
\usepackage{bbm}
\usepackage{bm}        
\usepackage{booktabs}  
\usepackage[mathscr]{eucal}
\usepackage{placeins}

\usepackage{algorithmic}
\usepackage{braket}
\usepackage{subcaption}
\usepackage{graphicx}
\usepackage[usenames,dvipsnames]{xcolor}

\usepackage{verbatim}

\usepackage{parskip} 

\usepackage{tikz}
\usetikzlibrary{quantikz}
\usepackage{hyperref}

\hypersetup{
  colorlinks=true,
  linkcolor=teal!60!black,
  citecolor=blue!60!black,
  urlcolor=violet!70!black
}

\begin{document}


\title{Coefficient-Decoupled Matrix Product Operators as an Interface to Linear-Combination-of-Unitaries Circuits}

\author{Younes Javanmard}
\email{younes.javanmard@itp.uni-hannover.de}
\affiliation{Institut f\"ur Theoretische Physik, Leibniz Universit\"at Hannover, Appelstra\ss e 2, 30167 Hannover, Germany}

\affiliation{Institut f\"ur Physik und Astronomie, Technische Universit\"at Berlin, Hardenbergstra\ss e 36, EW~7-1, 10623 Berlin, Germany}

\date{\today}

\begin{abstract}
We introduce a coefficient-decoupled matrix product operator (MPO) representation for Pauli-sum operators that separates reusable symbolic operator support from a tunable coefficient bridge across a fixed bipartition. This representation provides a direct interface to linear-combination-of-unitaries (LCU) circuits: the symbolic left/right dictionaries define a static \textsc{Select} oracle that is compiled once, while coefficient updates modify only a dynamic \textsc{Prep} oracle. As a proof of concept, we construct compact state-adapted Pauli pools by sampling Pauli strings from a pretrained matrix product state (MPS), pruning them to a fixed symbolic pool, optimizing only their coefficients, and transferring the resulting weights directly to the LCU interface. The resulting workflow provides a reusable classical-to-quantum compilation strategy in which the symbolic operator structure is compiled once, and subsequent updates are confined to a low-dimensional coefficient object.
\end{abstract}

\maketitle
\section{Introduction}

Matrix product operators (MPOs) provide a standard language for representing many-body operators in tensor-network methods, while linear-combination-of-unitaries (LCU) constructions provide a standard route for implementing structured operators in quantum algorithms \cite{NielsenChuang2010,BerryChildsCleveKothariSomma2015Taylor,BerryChildsKothari2015NearlyOptimal,LowChuang2019Qubitization,GilyenSuLowWiebe2019QSVT,ChakrabortyGilyenJeffery2019BlockEncoded, White1992DMRG,Schollwoeck2011DMRG,ZalatelMPO2015, PirvuMurgCiracVerstraete2010,CrosswhiteBacon2008,ChildsWiebe2012, LowChuang2019Qubitization, GilyenSuLowWiebe2019QSVT, NibbiBlockMPO2024, jiang2021chebyshev}. In both settings, an operator is often specified as a weighted sum of Pauli strings or other local operator strings. A common practical feature of such operators is that their symbolic support evolves much more slowly than their numerical coefficients: the same operator fragments reappear across parameter sweeps, optimization loops, variational updates, and data-driven fitting procedures, while only the associated weights change.

\begin{figure}[!t]
	\centering
	\includegraphics[width=0.83\columnwidth]{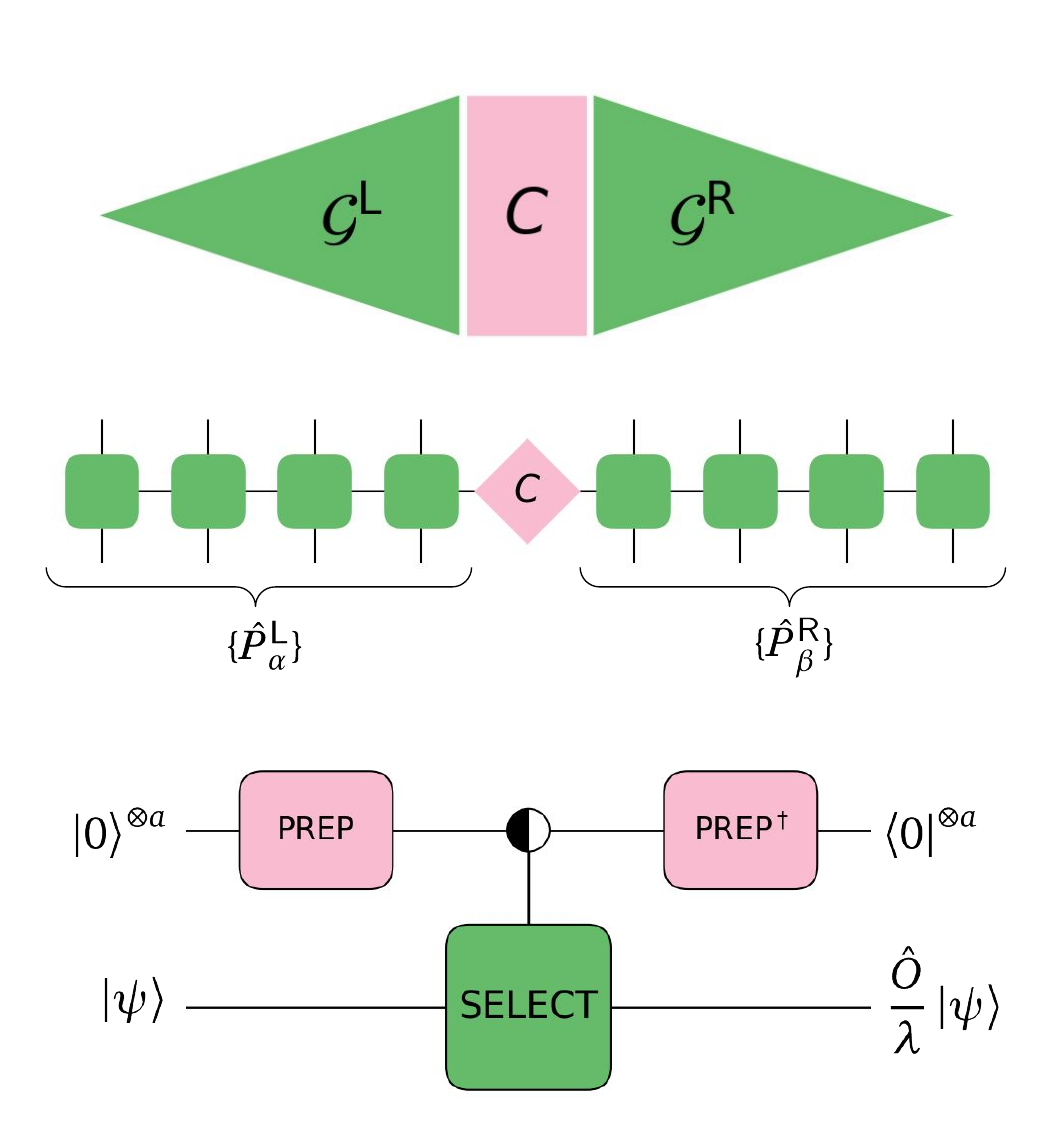}
	\caption[Coefficient-decoupled MPO and its LCU interpretation.]{Coefficient-decoupled MPO and its LCU interpretation. \emph{Top:} Coefficient-decoupled factorization across a fixed cut, \(\hat O=\sum_{(\alpha,\beta)\in \mathcal{E}} C_{\alpha\beta}\bigl(\hat P^{\mathsf L}_\alpha\otimes \hat P^{\mathsf R}_\beta\bigr)\), where the left/right graphs \(\mathcal{G}_{\mathsf L}\) and \(\mathcal{G}_{\mathsf R}\) encode the unique Pauli-fragment patterns \(\{\hat P^{\mathsf L}_\alpha\}\) and \(\{\hat P^{\mathsf R}_\beta\}\), and the typically sparse bridge matrix \(\mathbf{C}=[C_{\alpha\beta}]\) stores the coupling coefficients. \emph{Bottom:} LCU block encoding: \textsc{Prep} loads the coefficient distribution over an index register \(\protect\ket{\alpha}_{\mathsf L}\protect\ket{\beta}_{\mathsf R}\), \textsc{Select} applies the indexed unitaries \(\hat U_{\alpha\beta}=\hat P^{\mathsf L}_\alpha\otimes \hat P^{\mathsf R}_\beta\), and \textsc{Prep}$^\dagger$ uncomputes, yielding the \(\protect\ket{0}^{\otimes a}\)-block proportional to \((\hat O/\lambda)\protect\ket{\psi}\), with \(\lambda=\sum_{(\alpha,\beta)\in \mathcal{E}} |C_{\alpha\beta}|\). Here \(a\) denotes the number of ancilla qubits in the index register. For simplicity, sign and phase handling are omitted; negative or complex coefficients can be incorporated in the standard LCU manner.}
	\label{fig:fig1_mpo_diag}
\end{figure}

This observation motivates a compilation strategy in which symbolic structure and numerical coefficients are treated separately. Rather than rebuilding a full MPO whenever the coefficients change, we seek a representation in which the operator-string structure is compiled once into reusable symbolic components, while all coefficient updates are confined to a small numerical object. In this work, we formalize this idea through a coefficient-decoupled MPO representation. Across a fixed bipartition, the operator is expressed in terms of reusable left and right Pauli-fragment dictionaries together with a coefficient bridge that couples the two sides. The symbolic structure on the left and right is fixed, while all numerical freedom is isolated in the bridge.

This separation is especially useful because it exposes a direct interface to LCU circuits \cite{ChildsWiebe2012,LowChuang2019Qubitization,GilyenSuLowWiebe2019QSVT, rudolph2025pauli}. Given left and right fragment dictionaries \(\{\hat P^{\mathsf L}_\alpha\}\) and \(\{\hat P^{\mathsf R}_\beta\}\), each basis term
\(\hat U_{\alpha\beta}=\hat P^{\mathsf L}_\alpha\otimes \hat P^{\mathsf R}_\beta\)
is unitary, and the bridge coefficients define the corresponding LCU weights. In this mapping, the \textsc{Select} oracle depends only on the symbolic dictionaries and can therefore be compiled once and reused, whereas the \textsc{Prep} oracle depends on the bridge coefficients and is the only component that must be updated when the coefficients change. Coefficient-decoupled MPOs therefore provide a natural static/dynamic split for LCU compilation: the symbolic compilation is fixed, while coefficient loading remains dynamic.

Our second focus is a concrete workflow that exploits this interface. Instead of starting from a hand-designed Pauli pool, we construct a state-adapted pool by sampling Pauli strings from a pretrained matrix product state (MPS), for example, a high-quality DMRG reference state \cite{ LamiCollura2023, gu2025quantum, sheng2024td, li2023tencirchem, li2022fly, smith2024constant, malz2024preparation, fan2023quantum, schon2007sequential, schon2005sequential}. This biases the sampled operators toward directions that are relevant to the target state. In practice, naive sampling tends to overproduce diagonal strings composed only of \(I\) and \(Z\), which are often redundant for the intended variational span. We therefore retain only a small curated diagonal subset and prioritize off-diagonal strings containing \(X\) or \(Y\), yielding a compact Pauli pool that is better suited for generating correlations beyond simple diagonal reweighting. Once this symbolic pool is fixed, we optimize only the coefficients of a linear operator ansatz over the sampled strings. This coefficient-only workflow is conceptually close to state-adapted variational operator constructions while remaining compatible with tensor-network-pretrained quantum workflows and MPS-based variational ans\"atze \cite{GrimsleyEconomouBarnesMayhall2019, YJ_mps_ansatz2024, Javanmard2024PretrainedTNSearch, khan2023pre}. In this way, a pretrained classical tensor-network state is converted into a reusable LCU-ready operator scaffold whose symbolic structure is compiled once and whose subsequent updates are confined to coefficient loading.

The main contributions of this paper are as follows. First, we formulate a coefficient-decoupled MPO representation in which reusable left and right symbolic operator dictionaries are decoupled from a tunable coefficient bridge. Second, we show that this structure maps directly to LCU circuits with a static \textsc{Select} oracle and a dynamic \textsc{Prep} oracle. Third, we present a state-adapted workflow in which Pauli strings are sampled from a pretrained MPS, pruned to a compact pool, optimized only through their coefficients, and then transferred to the LCU representation.

The rest of the paper is organized as follows. In Sec.~\ref{sec:methods}, we review the operator-expansion setting and the LCU background relevant to our construction. In Sec.~\ref{sec:coeff-decoupleMPO}, we introduce the coefficient-decoupled MPO representation, the left/right symbolic factorization, and the coefficient bridge. In Sec.~\ref{sec:variational_coeff_decoupled_mpo}, we formulate variational ans\"atze in which only the bridge is optimized while the symbolic tensor-network structure is kept fixed. In Sec.~\ref{sec:app_chem}, we specialize the framework to electronic-structure Hamiltonians and state-adapted Pauli pools. In Sec.~\ref{sec:pauli_conditional_sampling_mps}, we describe the conditional sampling of Pauli strings from a reference MPS and its role in pool construction. In Sec.~\ref{sec:quantum_algorithm_interfaces}, we present the LCU interface and the associated static-\textsc{Select}/dynamic-\textsc{Prep} compilation strategy. In Sec.~\ref{sec:conclusion}, we conclude with limitations and outlook.

\section{Background}
\label{sec:methods}

Many physically relevant operators \(\hat O\) admit a finite expansion in a chosen local operator basis,
\begin{align}
\hat O=\sum_{l=1}^{K} c_l\,\hat B_l,
\label{eq:op_expansion_general}
\end{align}
where each \(\hat B_l\) is an operator string built from elementary on-site operators. For qubit systems, a natural choice is the Pauli basis \(\{I,X,Y,Z\}\), in which case the \(\hat B_l\) are Pauli strings. For fermionic problems, one may instead work directly in a fermionic operator basis generated by local building blocks such as \(\{I,\hat a_p,\hat a_p^\dagger,\hat n_p\}\), leading to strings such as \(\hat a_p^\dagger \hat a_q\) and \(\hat a_p^\dagger \hat a_q^\dagger \hat a_r \hat a_s\). Matrix product states and matrix product operators provide a standard tensor-network language for representing such structured many-body objects \cite{White1992DMRG,Schollwoeck2011DMRG,li2024optimal,ren2020general, guo2018matrix, cui2015variational, guo2022quantum}.

An operator \(\hat O\) is called \(k\)-local if each term \(\hat B_l\) acts nontrivially on at most \(k\) sites. In the qubit setting, where \(\hat B_l\) is a Pauli string
\[
\hat B_l=\bigotimes_{j=1}^{N}\hat\sigma_l^{(j)},
\qquad
\hat\sigma_l^{(j)}\in\{I,X,Y,Z\},
\]
this is equivalently characterized by the Pauli weight
\begin{align}
\mathrm{wt}(\hat B_l)=\left|\{j:\hat\sigma_l^{(j)}\neq I\}\right|,
\end{align}
so that \(\hat O\) is \(k\)-local if \(\mathrm{wt}(\hat B_l)\le k\) for all \(l\).

\subsection{Linear combination of unitaries}

In this work, we focus on the special case in which the expansion terms are unitary,
\begin{align}
\hat O=\sum_{l=1}^{K} c_l\,\hat U_l,
\label{eq:localO}
\end{align}
that is, the LCU form. In the qubit setting considered below, we take the unitary terms \(\hat U_l\) to be Pauli strings and write
\[
\hat P_l=\bigotimes_{j=1}^{N}\hat\sigma_l^{(j)},
\qquad
\hat\sigma_l^{(j)}\in\{I,X,Y,Z\}.
\]
Physical Hamiltonians are typically sparse in this basis, so only a small subset of the \(4^N\) Pauli strings appears with a nonzero coefficient. We restrict attention to the LCU setting because unitarity enables a direct compilation into standard quantum-algorithmic primitives such as \textsc{Prep}/\textsc{Select} block encodings. The central question addressed below is how to expose this structure directly at the MPO level in a form useful for both tensor-network optimization and LCU compilation.

\section{Coefficient-decoupled MPO construction}
\label{sec:coeff-decoupleMPO}

An operator sum of the form in Eq.~\eqref{eq:localO} can be converted into a matrix product operator (MPO) in several ways. One common route uses finite-state automata (FSAs), which are particularly effective for regular patterns such as Ising-type or Hubbard models \cite{CrosswhiteBacon2008}. A second route uses more general graph-based constructions, often referred to as tensor product diagrams (TPDs), which are well suited to complex and irregular Hamiltonians because repeated operator patterns can be deduplicated automatically, thereby controlling the MPO bond dimension \cite{ren2020general,HubigMcCullochSchollwoeck2017,Ran2020MPSCircuit,CakirMilbradtMendl2025,milbradt2026efficient, keller2015efficient, lukac2009quantum, kondacs1997power}.

In several variational and quantum-algorithmic settings, however, it is advantageous to separate the \emph{symbolic structure} of the operator from its \emph{numerical coefficients} at a fixed cut. This separation permits coefficient updates without rebuilding the underlying graph and, in the LCU setting, allows the same coefficient data to be transferred directly to the \textsc{Prep} stage without reconstructing the \textsc{Select} logic.

\subsection{Left/right factorization and coefficient bridge}
\label{sec:bridge}

To construct a symbolic MPO for \(\hat O\), we fix a bipartition of an \(N\)-qubit system into a left block
\[
\mathsf{L}=\{1,\dots,m\},
\qquad
\mathsf{R}=\{m+1,\dots,N\}.
\]
Any Pauli-string term \(\hat P_l\) factorizes across the cut as
\[
\hat P_l=\hat P_l^{\mathsf L}\otimes \hat P_l^{\mathsf R},
\]
where
\[
\hat P_l^{\mathsf L}=\bigotimes_{j\in\mathsf L}\hat\sigma_l^{(j)},
\qquad
\hat P_l^{\mathsf R}=\bigotimes_{j\in\mathsf R}\hat\sigma_l^{(j)}.
\]

For a weighted Pauli sum
\begin{equation}
\hat O=\sum_{l=1}^{K} c_l\,\hat P_l,
\end{equation}
we group repeated half-string patterns on each side of the cut. Let \(\mathcal{G}_{\mathsf L}\) and \(\mathcal{G}_{\mathsf R}\) denote the symbolic MPO graphs encoding the allowed Pauli fragments on the left and right, respectively. Let
\(\{\hat P_\alpha^{\mathsf L}\}_{\alpha\in\mathcal I_{\mathsf L}}\) and
\(\{\hat P_\beta^{\mathsf R}\}_{\beta\in\mathcal I_{\mathsf R}}\)
be the corresponding sets of unique left and right Pauli fragments. Then \(\hat O\) can be rewritten in the coefficient-decoupled form
\begin{align}
\hat O
=
\sum_{(\alpha,\beta)\in \mathcal E}
C_{\alpha\beta}\,
\bigl(\hat P_\alpha^{\mathsf L}\otimes \hat P_\beta^{\mathsf R}\bigr),
\label{eq:coeff_bridge_decomp}
\end{align}
where \(\mathcal E\subseteq \mathcal I_{\mathsf L}\times\mathcal I_{\mathsf R}\) is the active edge set induced by the symbolic MPO graphs. The matrix \(\mathbf C=[C_{\alpha\beta}]\), typically sparse, is the coefficient bridge that stores the numerical weights coupling the left and right fragments. Thus, the symbolic operator support is fixed by \(\mathcal G_{\mathsf L}\) and \(\mathcal G_{\mathsf R}\), while all numerical freedom is isolated in \(\mathbf C\).

At the MPO level, Eq.~\eqref{eq:coeff_bridge_decomp} may be viewed schematically as
\begin{align}
\hat O
=
\sum_{(\alpha,\beta)\in \mathcal E}
\mathcal W^{[\mathsf L]}_\alpha\,
C_{\alpha\beta}\,
\mathcal W^{[\mathsf R]}_\beta,
\label{eq:coeff_bridge_mpo}
\end{align}
where \(\mathcal W^{[\mathsf L]}_\alpha\) and \(\mathcal W^{[\mathsf R]}_\beta\) denote fixed symbolic MPO realizations of the left and right fragments. In this way, the symbolic MPO structure is compiled once, while coefficient updates are confined to the bridge.

In variational settings, we replace the fixed bridge \(\mathbf C\) by a trainable bridge \(\mathbf S_g(\vec\theta)\), whose entries \(s_{\alpha\beta}(\vec\theta)\) are determined by the parameter vector \(\vec\theta\). The resulting parameterized operator is interpreted as a generator
\begin{equation}
\hat G(\vec\theta)
=
\sum_{(\alpha,\beta)\in \mathcal E}
s_{\alpha\beta}(\vec\theta)\,
\bigl(\hat P_\alpha^{\mathsf L}\otimes \hat P_\beta^{\mathsf R}\bigr),
\label{eq:G_coeff_decoupled}
\end{equation}
so that all variational freedom is confined to the coefficient bridge \(\mathbf S_g(\vec\theta)=\bigl[s_{\alpha\beta}(\vec\theta)\bigr]\).

Equation~\eqref{eq:G_coeff_decoupled} therefore defines a parameterized MPO ansatz in which one optimizes only the bridge coefficients while the left and right symbolic graphs remain fixed.

\subsubsection{Construction algorithm}
\label{subsubsec:construction_algorithm}

At a fixed cut \(m\), each Pauli string is decomposed into a left fragment on \(\mathsf L=\{1,\dots,m\}\) and a right fragment on \(\mathsf R=\{m+1,\dots,N\}\). Repeated half-string patterns are then deduplicated into symbolic left and right graphs, \(\mathcal G_{\mathsf L}\) and \(\mathcal G_{\mathsf R}\), while the numerical coefficients are collected into a sparse bridge matrix \(\mathbf C\) indexed by fragment pairs \((\alpha,\beta)\). The explicit symbolic-compilation procedure is given in Algorithm~\ref{alg:lr_bridge} in Appendix~\ref{app:symbolic-graphs-bridge}.

The coefficient bridge \(\mathbf C=[C_{\alpha\beta}]\) also provides a natural compression point for the coefficient-decoupled MPO. When the left and right fragment bases are orthonormal with respect to the Hilbert–Schmidt inner product, a low-rank approximation of \(\mathbf C\) directly yields a reduced effective bond dimension across the chosen cut. In this sense, the bridge serves both as a compact carrier of numerical coefficients and as a diagnostic of the operator complexity across the left/right bipartition. Standard canonical-form MPO compression and its relation to this bridge-based viewpoint are summarized in Appendix~\ref{app:mpo_compression}.
\begin{figure}[th!]
  \centering
  \begin{subfigure}[t]{\columnwidth}
  \caption{}
    \centering
    \includegraphics[width=\columnwidth]{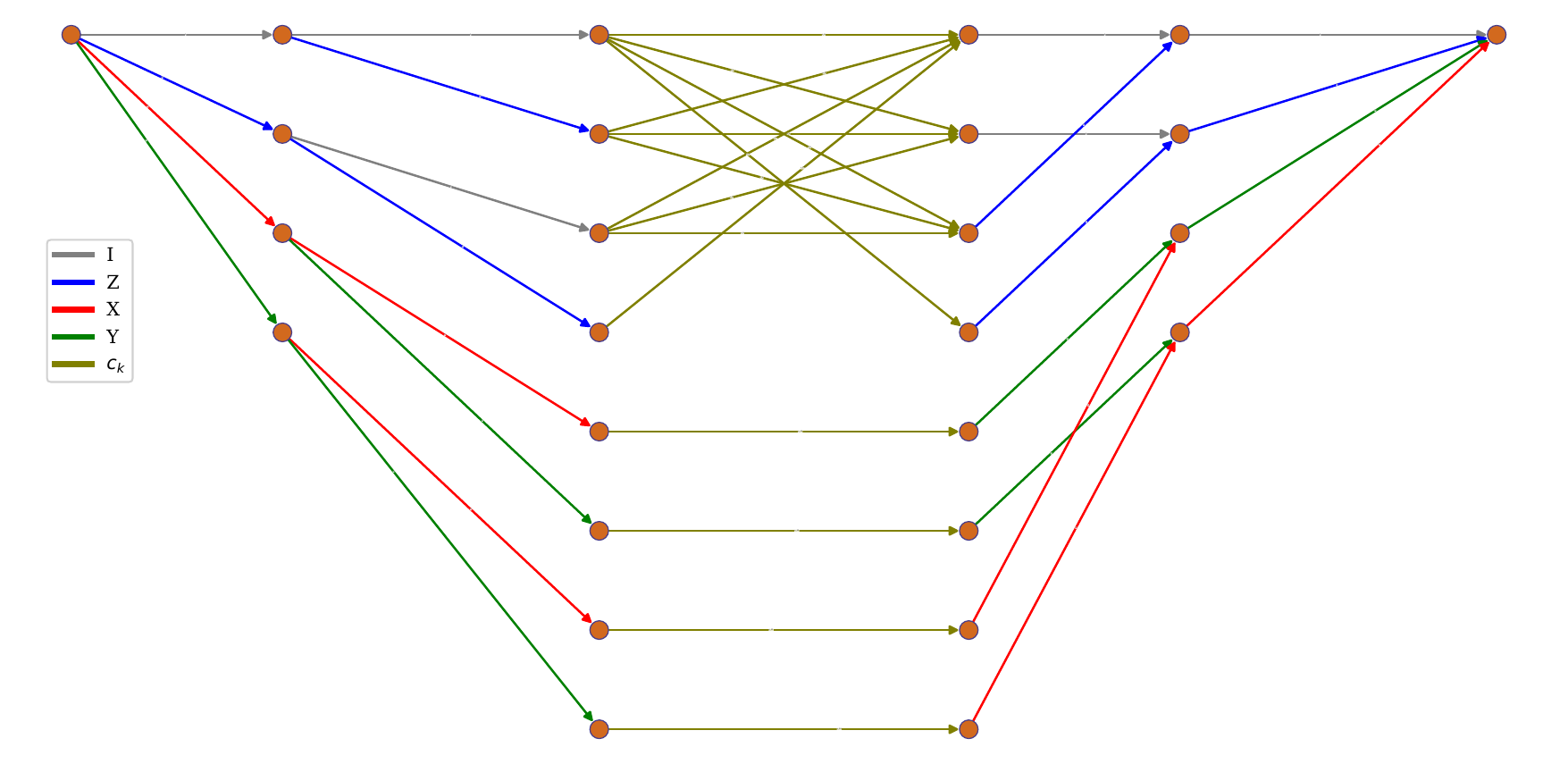}
  \end{subfigure}\hfill
  \begin{subfigure}[t]{\columnwidth}
  \caption{}
    \centering
    \includegraphics[width=\columnwidth]{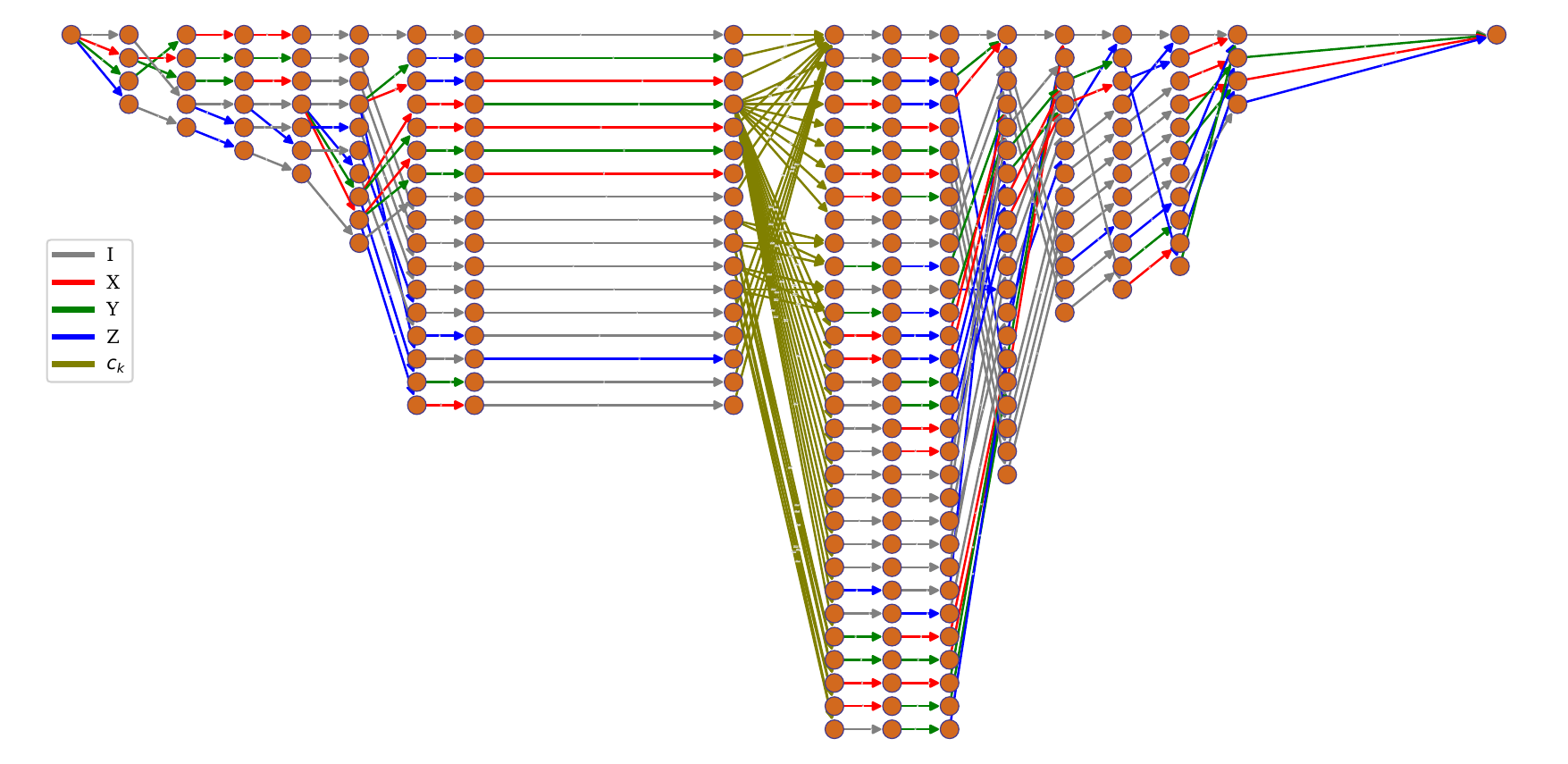}
  \end{subfigure}
  \caption{Examples of coefficient-decoupled MPO constructions with an explicit bridge. (a) The \(\mathrm{H}_2\) electronic Hamiltonian in the \textit{sto-3g} basis, mapped to qubits and written as a Pauli sum \(\hat H=\sum_i c_i(R)\hat P_i\). As the bond length \(R\) varies, the coefficients change while the Pauli support, and hence the symbolic graph structure, remain fixed. The graph separates into a diagonal \(I/Z\) sector and an off-diagonal sector containing strings with at least one \(X\) or \(Y\), reflecting symmetry-resolved structure that can persist in larger molecules with additional symmetries \cite{singh2011tensor,singh2012tensor}. (b) The \(\mathrm{H}_8\) chain Hamiltonian in the \textit{sto-3g} basis with 8 orbitals (16 qubits), shown using a randomly selected subset of 50 Pauli terms for visualization. In both panels, the left and right subgraphs encode deduplicated Pauli-string fragments on either side of the cut, while the bridge stores the coefficients coupling the two symbolic graphs.}
  \label{fig:h2_h8_mpo}
\end{figure}

\begin{figure*}[t]
\centering
\includegraphics[width=\textwidth]{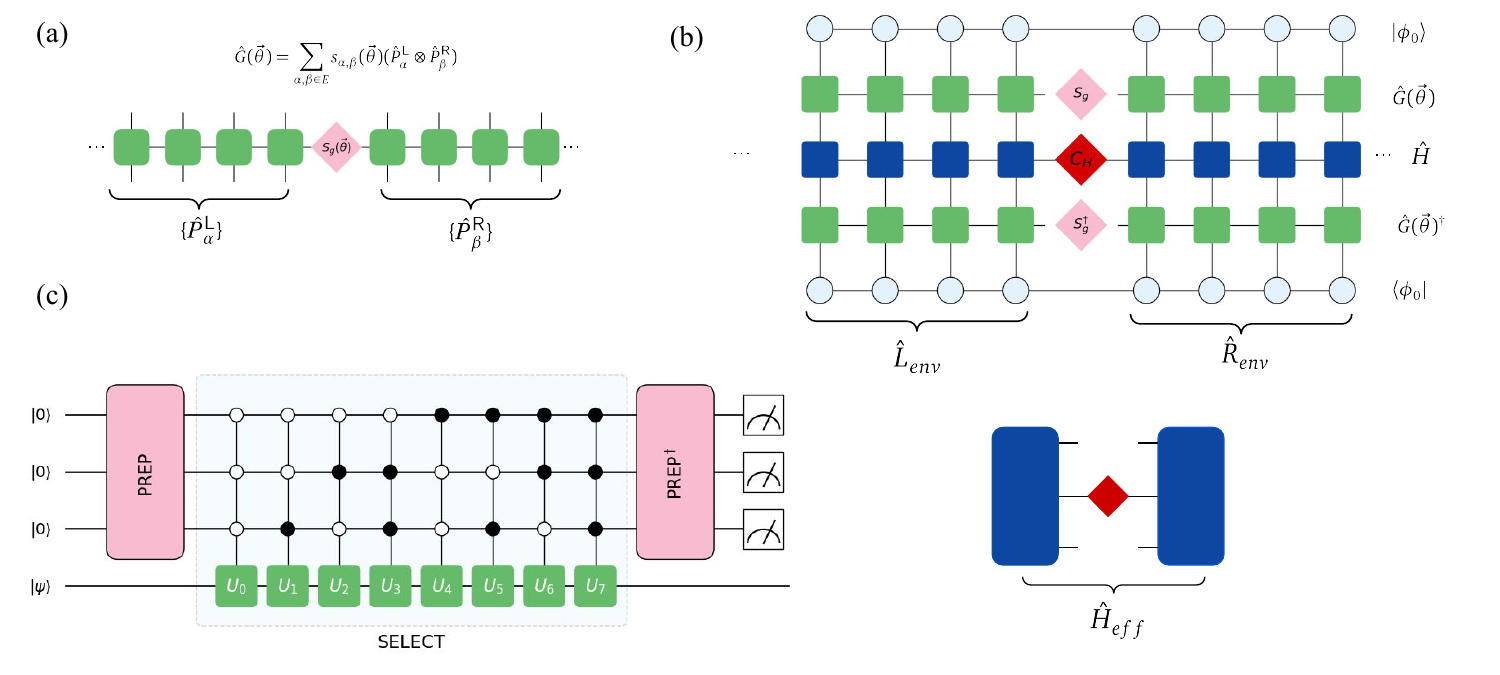}
\caption{Coefficient-decoupled MPOs and their LCU oracle interface. (a) \emph{Coefficient-decoupled generator.} Across a bipartition \(\mathsf{L}\cup\mathsf{R}\), the generator \(\hat G(\vec\theta)=\sum_{(\alpha,\beta)\in\mathcal E}s_{\alpha\beta}(\vec\theta)\bigl(\hat P^{\mathsf L}_\alpha\otimes \hat P^{\mathsf R}_\beta\bigr)\) is specified by fixed left/right Pauli-fragment dictionaries and a tunable bridge \(\mathbf S_g(\vec\theta)\). (b) \emph{Effective environment contractions.} Contracting \(\bra{\phi_0}\hat G^\dagger(\vec\theta)\hat H\hat G(\vec\theta)\ket{\phi_0}\) yields left and right environments and an effective problem at the cut, such as \(H_{\rm eff}\) and \(N_{\rm eff}\), acting only on the bridge indices. (c) \emph{Static-\textsc{Select}/dynamic-\textsc{Prep} interface.} The bridge coefficients define a dynamic \textsc{Prep} oracle on the index register \((\alpha,\beta)\), while a static \textsc{Select} oracle applies \(\hat U_{\alpha\beta}=\hat P^{\mathsf L}_\alpha\otimes \hat P^{\mathsf R}_\beta\). The resulting circuit block encodes \(\hat G(\vec\theta)/\Lambda(\vec\theta)\), where \(\Lambda(\vec\theta)=\sum_{(\alpha,\beta)\in\mathcal E}|s_{\alpha\beta}(\vec\theta)|\). Signs and phases may be absorbed into the selected unitaries in the standard LCU manner.}
\label{fig:MPO-LCU}
\end{figure*}
\section{Variational ansatz with a coefficient-decoupled MPO}
\label{sec:variational_coeff_decoupled_mpo}

A central application of the coefficient-decoupled MPO construction is the variational approximation of low-energy states of a target Hamiltonian \(\hat H\), with electronic-structure Hamiltonians providing a primary example. We use the Rayleigh–Ritz variational principle to minimize the energy over a chosen variational class. For context, standard DMRG optimizes over the manifold of matrix product states (MPS) with fixed bond dimension \(\chi\) \cite{White1992DMRG,Schollwoeck2011DMRG}.

\paragraph{Coefficient-decoupled variational class.}
Here we instead consider an \emph{operator-generated} ansatz tailored to the coefficient-decoupled MPO structure. Rather than optimizing all local tensors, we keep the left and right symbolic tensor-network structures fixed and optimize only the coefficient bridge at a chosen cut.

Starting from a reference state \(\ket{\phi_0}\), typically a mean-field or product state, we apply a generator \(\hat G(\mathbf S_g)\) of the form in Eq.~\eqref{eq:G_coeff_decoupled}. Because the tensor network depends linearly on the bridge entries, flattening the bridge index pair \((\alpha,\beta)\) into a single index \(j\) and writing
\[
x=\mathrm{vec}(\mathbf S_g)\in\mathbb C^{d_{\mathrm{br}}},
\qquad
d_{\mathrm{br}}=|\mathcal I_{\mathsf L}|\,|\mathcal I_{\mathsf R}|,
\]
gives the linear trial-state expansion
\begin{equation}
\ket{\psi(\mathbf S_g)}
\equiv
\ket{\psi(x)}
=
\sum_{j=1}^{d_{\mathrm{br}}} x_j\,\ket{\varphi_j},
\label{eq:psi_linear_expansion}
\end{equation}
where \(\{\ket{\varphi_j}\}\) is obtained by contracting all fixed tensors while selecting a basis element on the bridge index pair. Equivalently, the same variational family may be viewed as a linear operator ansatz over a fixed unitary dictionary indexed by \((\alpha,\beta)\).

The ground-state search, therefore, reduces to
\begin{equation}
\begin{aligned}
\ket{\psi_{\rm gs}}
&\sim
\arg\min_{\mathbf S_g}
\frac{
\bra{\phi_0}\hat G^\dagger(\mathbf S_g)\,\hat H\,\hat G(\mathbf S_g)\ket{\phi_0}
}{
\bra{\phi_0}\hat G^\dagger(\mathbf S_g)\hat G(\mathbf S_g)\ket{\phi_0}
}
\\
&=
\arg\min_{x}
\frac{x^\dagger H_{\rm eff}x}{x^\dagger N_{\rm eff}x},
\end{aligned}
\label{eq:bridge_variational_ritz}
\end{equation}
with effective matrices
\begin{equation}
(H_{\rm eff})_{jk}=\bra{\varphi_j}\hat H\ket{\varphi_k},
\qquad
(N_{\rm eff})_{jk}=\braket{\varphi_j|\varphi_k}.
\label{eq:heff_neff_defs}
\end{equation}
Because the ansatz is linear in \(x\), the effective optimization problem can be handled without explicitly assembling \(H_{\rm eff}\) and \(N_{\rm eff}\); one only needs their actions on trial vectors.

\subsubsection{Ground-state optimization}
\label{sec:ground_state_optimization}

Stationarity of Eq.~\eqref{eq:bridge_variational_ritz} yields the generalized eigenvalue problem
\begin{equation}
H_{\rm eff}x=\varepsilon\,N_{\rm eff}x,
\label{eq:generalized_eig}
\end{equation}
whose smallest eigenpair gives the optimal bridge coefficients \(x\), and hence the optimal bridge matrix \(\mathbf S_g\). In practice, Eq.~\eqref{eq:generalized_eig} can be solved with iterative eigensolvers such as locally optimal block preconditioned conjugate gradient (LOBPCG) \cite{Knyazev2001LOBPCG,Saad2011LargeEigen,SunMilbradtKnechtKumarMendl2025}, requiring only the matrix-vector actions \(x\mapsto H_{\rm eff}x\) and \(x\mapsto N_{\rm eff}x\), implemented directly through tensor-network contractions.

\paragraph{Fixed symbolic span and activated terms.}
Starting from an initial Pauli pool, one may split it into left and right symbolic dictionaries, or equivalently into symbolic subgraphs \(\mathcal G_{\mathsf L}\) and \(\mathcal G_{\mathsf R}\), and then optimize only the bridge coefficients \(\mathbf S_g=[s_{\alpha\beta}]\). During the optimization, entries that were initially zero may become nonzero, thereby activating additional Pauli products in \(\hat G\). This does not alter the underlying operator dictionary: the fragment sets \(\{\hat P^{\mathsf L}_\alpha\}\) and \(\{\hat P^{\mathsf R}_\beta\}\) remain fixed, so the map
\[
(\alpha,\beta)\mapsto \hat U_{\alpha\beta}
=
\hat P^{\mathsf L}_\alpha\otimes \hat P^{\mathsf R}_\beta
\]
is unchanged. New terms appear only because the optimizer explores a larger portion of the fixed linear span generated by the precompiled symbolic dictionaries.

\section{Applications: electronic-structure specialization and state-adapted Pauli pools}
\label{sec:app_chem}

We now specialize the coefficient-decoupled MPO framework and its LCU interface to electronic structure \cite{McArdleEndoAspuruGuzikBenjaminYuan2020,Fulde2019Wavefunctions,fulde1995electron,kochman1998jordan}. Starting from the second-quantized molecular Hamiltonian and applying a standard fermion-to-qubit mapping, we obtain a qubit operator of the form
\begin{equation}
\hat H=\sum_i c_i\,\hat P_i,
\qquad
\hat P_i\in\{I,X,Y,Z\}^{\otimes N},
\qquad
c_i\in\mathbb R,
\label{eq:H_pauli}
\end{equation}
where the coefficients \(c_i\) are computed classically from the molecular integrals. Details of the fermionic Hamiltonian and the fermion-to-qubit mapping are given in Appendix~\ref{app:chem_prelim}.

\subsection{State-adapted Pauli pools from a pretrained matrix product state}
\label{sec:state_adapted_pool}

Rather than constructing a Pauli pool from a predetermined set of excitation operators, we build a compact \emph{state-adapted} pool by sampling Pauli strings from a high-quality reference matrix product state (MPS), for example, a DMRG ground-state approximation \cite{White1992DMRG,Schollwoeck2011DMRG,LamiCollura2023}. Concretely, we first compute an MPS approximation \(\ket{\psi_{\rm ref}}\) to the target state and then sample Pauli strings \(P\) from a distribution \(\Pi(P)\) induced by \(\ket{\psi_{\rm ref}}\); see Sec.~\ref{sec:pauli_conditional_sampling_mps}. This biases the sampled operators toward directions relevant to the target state.

In practice, naive sampling is often dominated by diagonal Pauli strings containing only \(I\) and \(Z\). Since these mainly generate density-like reweightings and energy shifts, we do not retain this sector uniformly. Instead, we keep only a small curated subset of purely diagonal sampled strings and prioritize strings containing at least one \(X\) or \(Y\). Accordingly, we define
\[
\mathcal U_{XY}
=
\{\,P:\; P \text{ is sampled and contains at least one } X \text{ or } Y\,\},
\]
and let \(\mathcal U_{IZ}\) denote a small retained subset of sampled diagonal strings containing only \(I\) and \(Z\). The final curated pool is then
\[
\mathcal U=\mathcal U_{IZ}\cup\mathcal U_{XY},
\qquad
P_{\mathrm{pool}}:=|\mathcal U|.
\]
In this way, \(\mathcal U_{XY}\) supplies the noncommuting directions needed to generate correlations beyond mean field, while \(\mathcal U_{IZ}\) provides a minimal diagonal dressing without overparameterizing a largely redundant sector.

\subsection{Coefficient-only training on a fixed symbolic pool}
\label{sec:coeff_only_training}

Given a fixed curated pool \(\mathcal U=\{P_j\}_{j=1}^{P_{\mathrm{pool}}}\), we define the linear operator ansatz
\begin{equation}
\hat G(\vec\alpha)=\sum_{j=1}^{P_{\mathrm{pool}}}\alpha_j\,P_j,
\label{eq:G_lcu_global_main}
\end{equation}
and the associated trial state
\begin{equation}
\ket{\psi(\vec\alpha)}=\hat G(\vec\alpha)\ket{\Phi_0},
\end{equation}
where \(\ket{\Phi_0}\) is a classical reference state, such as Hartree–Fock. Because \(\hat G(\vec\alpha)\) is linear in the coefficients \(\vec\alpha\), the optimization reduces to a generalized Ritz problem in the span of the states
\begin{equation}
\ket{\varphi_j}:=P_j\ket{\Phi_0}.
\end{equation}
The corresponding effective matrices are
\begin{equation}
(H_{\rm eff})_{jk}
=
\bra{\Phi_0}P_j^\dagger \hat H P_k\ket{\Phi_0},
\qquad
(N_{\rm eff})_{jk}
=
\bra{\Phi_0}P_j^\dagger P_k\ket{\Phi_0},
\label{eq:Heff_Neff}
\end{equation}
and solving the generalized eigenvalue problem for \((H_{\rm eff},N_{\rm eff})\) yields the trained coefficient vector \(\vec\alpha\).

This flat-pool ansatz is equivalent to the bridge formulation of Sec.~\ref{sec:variational_coeff_decoupled_mpo}. Once the pool \(\mathcal U\) is organized into left/right fragment dictionaries, each index \(j\) may be viewed as a flattened version of a bridge index pair \((\alpha,\beta)\), with
\begin{equation}
P_j \equiv U_{\alpha\beta}
=
\hat P^{\mathsf L}_\alpha\otimes \hat P^{\mathsf R}_\beta.
\end{equation}
Thus, coefficient-only training over the fixed pool \(\mathcal U\) realizes the same variational family as bridge optimization over a fixed symbolic span.

Figure~\ref{fig:excitation_vs_samples} shows the optimized energy as a function of \(N_{\rm samples}\), the number of Pauli strings drawn from the pretrained MPS before curation, with the Hartree--Fock state \(\ket{\Phi_{\mathrm{HF}}}\) used as the initial reference. For each value of \(N_{\rm samples}\), we construct a curated pool by retaining strings containing at least one \(X\) or \(Y\), together with only a small retained subset of purely diagonal \(I/Z\)-type strings. For bond dimension \(\chi=8\), optimizing the coefficients within this sampled pool already lowers the variational energy. As the bond dimension increases and the pretrained MPS approaches the true ground state more closely, larger values of \(N_{\rm samples}\) are needed to approach the DMRG benchmark. In that regime, the optimized coefficient-decoupled generator provides a good approximation to the DMRG ground-state energy \cite{YJ_mps_ansatz2024,Javanmard2024PretrainedTNSearch}.

\section{Conditional sampling of Pauli strings from a matrix product state}
\label{sec:pauli_conditional_sampling_mps}

We next describe how to sample Pauli strings
\(
P=\sigma_1\otimes\cdots\otimes\sigma_N
\),
with
\(
\sigma_j\in\{I,X,Y,Z\}
\),
from a target distribution \(\Pi(P)\) defined by a reference matrix product state \(\ket{\psi}\). Matrix product states provide a natural tensor-network representation for one-dimensional many-body wave functions and their numerical approximations \cite{White1992DMRG,Schollwoeck2011DMRG,javanmard_sharp_2018}. The central idea is to generate \(P\) sequentially by factorizing the target probability into conditional probabilities and updating a left environment after each local choice. This yields \emph{perfect sampling}, i.e., each Pauli string is drawn directly from the desired distribution without any Markov-chain equilibration \cite{LamiCollura2023}.

Let \(\Pi(P)\) be the Pauli-string distribution of interest. In the present application, \(\Pi\) is chosen so that strings with larger probabilities are more relevant to the reference state \(\ket{\psi}\), and hence more useful for building a state-adapted pool. The sampling strategy is based on the chain rule,
\begin{equation}
\Pi(\sigma_1,\ldots,\sigma_N)
=
\pi(\sigma_1)\prod_{j=2}^{N}\pi(\sigma_j\mid \sigma_1,\ldots,\sigma_{j-1}),
\label{eq:chain_rule}
\end{equation}
where \(\pi(\sigma_j\mid \sigma_{<j})\) denotes the conditional probability for the local Pauli operator at site \(j\), given the previously sampled prefix \(\sigma_{<j}\equiv(\sigma_1,\ldots,\sigma_{j-1})\). Sampling therefore proceeds from left to right: at each site, one computes the four conditional probabilities for \(I,X,Y,Z\), draws the local Pauli operator from this categorical distribution, and updates a compact environment object that carries all the information needed for the next step.

\begin{figure*}[th!]
  \centering
  \begin{subfigure}[t]{0.5\textwidth}
   \captionsetup{justification=raggedright,singlelinecheck=false}
   \caption{}
    \centering
    \includegraphics[width=0.9\columnwidth]{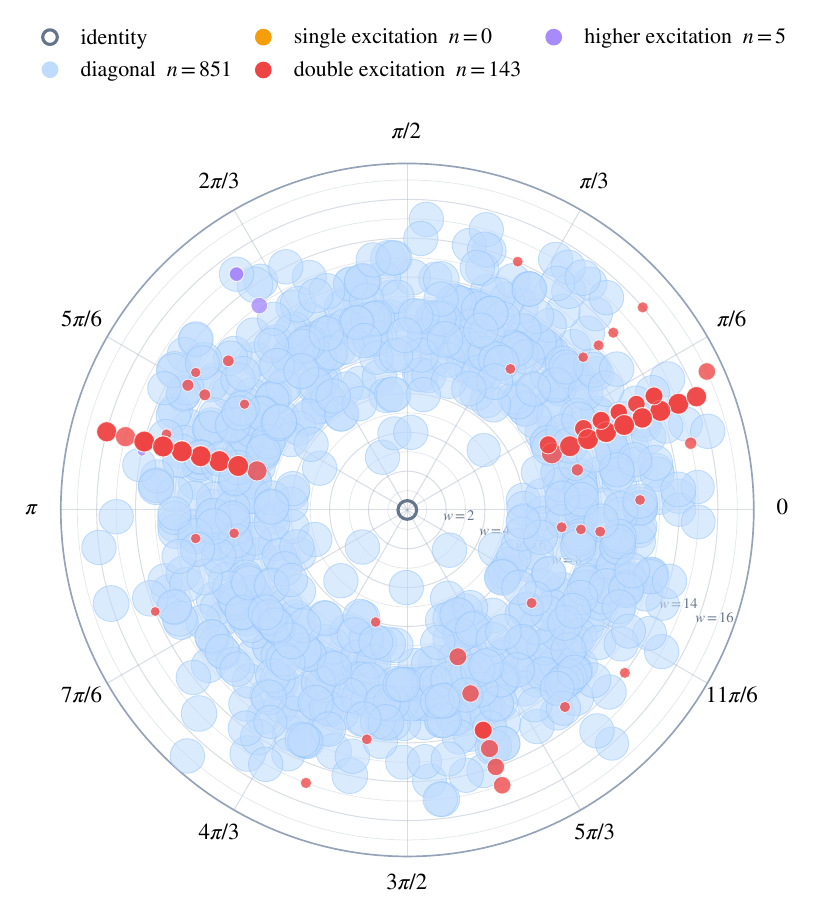}
  \end{subfigure}\hfill
  \begin{subfigure}[t]{0.5\textwidth}
   \captionsetup{justification=raggedright,singlelinecheck=false}
   \caption{}
    \centering
    \includegraphics[width=0.9\columnwidth]{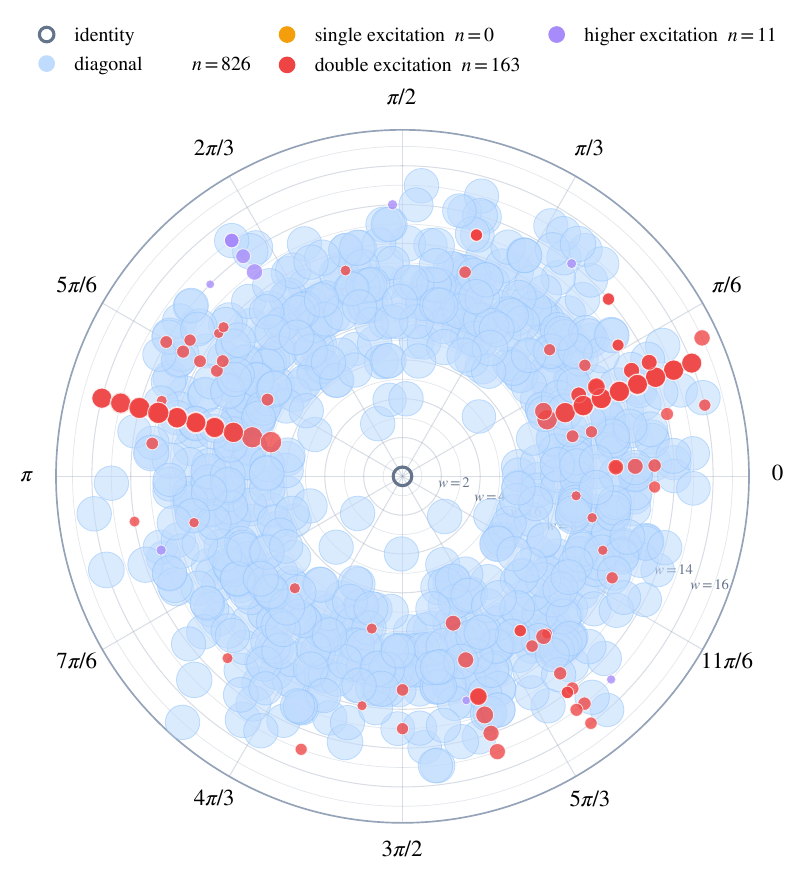}
  \end{subfigure}
  \caption{Polar map of Pauli strings sampled from pretrained MPSs for \(\mathrm{C}_2\mathrm{H}_4\), classified by excitation order: (a) \(\chi=8\) and (b) \(\chi=32\). The radial coordinate is the normalized Pauli weight \(r(P)=w(P)/w_{\max}\), where \(w(P)=|\{i:P_i\neq I\}|\). The angle \(\theta(P)\in[0,2\pi)\) is assigned through a bitmask-based hash of the operator support and is used only to distribute different support patterns visually around the circle; here \(b(P)=\sum_i \mathbf{1}[P_i\in\mathcal A]\,2^{i-1}\), with \(\mathcal A=\{X,Y\}\) for excitation strings and \(\mathcal A=\{Z\}\) for diagonal strings. Dot area scales as \(\sqrt{\Pi(P)}\), so more probable strings appear larger, and the hollow circle at the origin denotes \(I^{\otimes N}\). In both panels, the sampled distribution is dominated by diagonal strings, while the off-diagonal weight is concentrated mainly in the double-excitation sector; higher-excitation strings are comparatively rare.}
\label{fig:pauli_polar}
\end{figure*}

\begin{figure}[!htbp]
    \centering
    \includegraphics[width=\columnwidth]{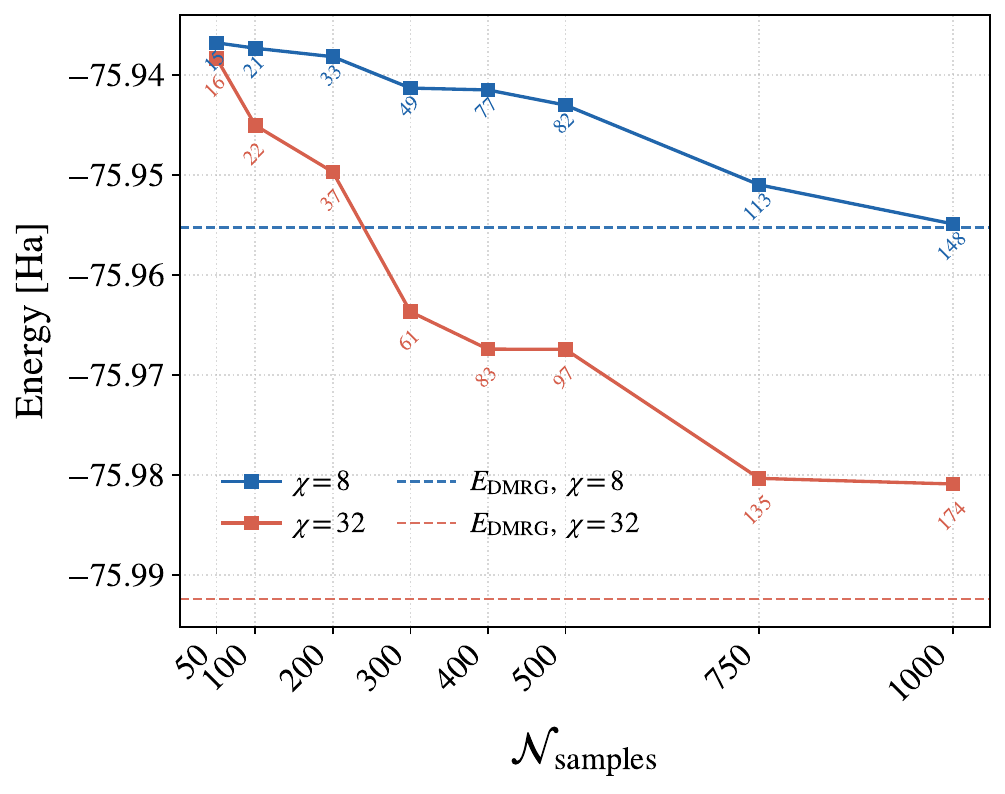}
    \caption{Optimized energy of the molecule \(C_2H_4\) in the \textit{sto-3g} basis with $10$ orbitals ($20$ qubits) with the ansatz \(\ket{\psi(\vec\alpha)}=\hat G(\vec\alpha)\ket{\Phi_0}\) as a function of the number \(N_{\rm samples}\) of Pauli strings sampled from a pretrained MPS, for reference bond dimensions \(\chi=8\) and \(\chi=32\) corresponding to Figure~\ref{fig:pauli_polar}. For each value of \(N_{\rm samples}\), a curated pool is constructed by retaining strings containing at least one \(X\) or \(Y\), together with only a small retained subset of purely diagonal \(I/Z\)-type strings. The dashed horizontal lines show the corresponding DMRG benchmark energies. Increasing \(N_{\rm samples}\) systematically lowers the variational energy, but the result saturates above the DMRG benchmark, indicating convergence within the sampled operator span rather than to the full DMRG manifold. The \(\chi=32\) reference captures richer correlations and therefore requires larger values of \(\mathcal{N}_{\rm samples}\) to approach its DMRG energy.}
    \label{fig:excitation_vs_samples}
\end{figure}

Assume that \(\ket{\psi}\) is represented as an MPS,
\begin{equation}
\ket{\psi}
=
\sum_{s_1,\ldots,s_N}
A^{s_1}A^{s_2}\cdots A^{s_N}\ket{s_1\cdots s_N},
\end{equation}
with bond dimension \(\chi\), in a canonical gauge satisfying the normalization condition
\begin{equation}
\sum_{s_j} A^{s_j\dagger}A^{s_j}=\mathrm{I}.
\end{equation}

The conditional probabilities are obtained from local contractions involving the tensor \(A^{s_j}\), its conjugate, the candidate local Pauli operator, and a left environment matrix \(E_{j-1}\) that summarizes the previously sampled prefix. More explicitly, at site \(j\) one forms four unnormalized weights
\begin{equation}
\widetilde{\pi}_j(\alpha),
\qquad
\alpha\in\{I,X,Y,Z\},
\end{equation}
by contracting the local network with the current environment \(E_{j-1}\) and then normalizing it accordingly
\begin{equation}
\pi(\alpha\mid \sigma_{<j})
=
\frac{\widetilde{\pi}_j(\alpha)}
{\sum_{\beta\in\{I,X,Y,Z\}}\widetilde{\pi}_j(\beta)}.
\end{equation}
After \(\sigma_j\) is sampled, the environment is updated by the corresponding local contraction and passed to the next site. Repeating this procedure from \(j=1\) to \(N\) produces an exact sample from \(\Pi(P)\).

In practice, perfect sampling is implemented by sweeping from left to right through the canonical MPS, evaluating the conditional probabilities for \(I,X,Y,Z\) at each site, sampling the local Pauli operator, and updating the corresponding left environment; Appendix~\ref{app:perfect_sampling} gives the full algorithm. For an MPS with bond dimension \(\chi\), each local probability evaluation and environment update involves contractions of \(\chi\times\chi\) objects and therefore costs \(O(\chi^3)\), assuming dense bond spaces. Generating one Pauli string thus costs \(O(N\chi^3)\), and generating \(\mathcal N_{\rm samp}\) independent samples costs \(O(\mathcal N_{\rm samp}\,N\,\chi^3)\).

Figure~\ref{fig:pauli_polar} visualizes the Pauli strings sampled from the pretrained MPS ground state of the \(\mathrm{C}_2\mathrm{H}_4\) molecule with $10$ orbitals ($20$ qubits) in a polar representation, grouped by excitation order. The radial coordinate encodes the normalized Pauli weight, while the angular coordinate distinguishes different operator-support patterns through the bitmask hash. For both \(\chi=8\) and \(\chi=32\), the sampled distribution is dominated by diagonal strings, whereas the off-diagonal weight is concentrated primarily in the double-excitation sector; higher excitations occur only sparsely. Increasing the bond dimension from \(\chi=8\) to \(\chi=32\) slightly broadens the off-diagonal distribution and increases the presence of higher-excitation strings, indicating that the higher-quality reference state captures richer correlations and therefore supports a more diverse state-adapted operator pool.

Once the sampled pool \(\mathcal U\) is fixed and the coefficients are optimized, the ansatz is naturally transferred to the coefficient-decoupled MPO representation. The symbolic support is compiled once, while later updates affect only the coefficients. This makes the construction well suited for an LCU implementation. In the next section, we show how this leads to a static-\textsc{Select}/dynamic-\textsc{Prep} decomposition.
\section{Quantum algorithm interface}
\label{sec:quantum_algorithm_interfaces}

\subsection{LCU interface for coefficient-decoupled MPOs}

Given the coefficient-decoupled MPO generator \(\hat G(\vec\theta)\) from Sec.~\ref{sec:coeff-decoupleMPO}, the left and right Pauli-fragment dictionaries \(\{\hat P^{\mathsf L}_\alpha\}\) and \(\{\hat P^{\mathsf R}_\beta\}\) act on disjoint subsystems. Hence, each basis term
\begin{equation}
\hat U_{\alpha\beta}
:=
\hat P^{\mathsf L}_\alpha\otimes \hat P^{\mathsf R}_\beta
\end{equation}
is unitary.

This directly induces an LCU implementation. The fixed symbolic structure of the operator determines a reusable \textsc{Select} oracle, while the numerical bridge coefficients determine a dynamic \textsc{Prep} oracle. Thus, once a Pauli pool has been sampled, curated, and trained, the same symbolic support can be reused across coefficient updates without recompiling the selection logic. The resulting workflow is
\begin{align*}
&\text{(i) sample and curate } \mathcal U,\\
&\text{(ii) solve for } \vec\alpha \text{ or } \mathbf S_g(\vec\theta),\\
&\text{(iii) update \textsc{Prep} and reuse \textsc{Select}.}
\end{align*}
Figure~\ref{fig:MPO-LCU} summarizes this static-\textsc{Select}/dynamic-\textsc{Prep} interface.

Let \(\mathcal E\) denote the active edge set induced by the symbolic MPO graphs, and define the \(\ell_1\) normalization
\begin{equation}
\Lambda(\vec\theta)
:=
\sum_{(\alpha,\beta)\in\mathcal E}
|s_{\alpha\beta}(\vec\theta)|,
\qquad
p_{\alpha\beta}(\vec\theta)
:=
\frac{|s_{\alpha\beta}(\vec\theta)|}{\Lambda(\vec\theta)}.
\label{eq:lcu_probs}
\end{equation}
Here \(a\) denotes the number of ancilla qubits in the index register used to encode the pair \((\alpha,\beta)\). The corresponding \textsc{Prep} oracle loads the coefficient magnitudes into amplitudes on that register:
\begin{equation}
\textsc{Prep}(\vec\theta):\quad
\ket{0}^{\otimes a}
\mapsto
\sum_{(\alpha,\beta)\in\mathcal E}
\sqrt{p_{\alpha\beta}(\vec\theta)}\,
\ket{\alpha, \beta}.
\label{eq:prep}
\end{equation}
The \textsc{Select} oracle applies the indexed unitary basis term,
\begin{equation}
\textsc{Select}:\quad
\ket{\alpha,\beta}\ket{\psi}
\mapsto
\ket{\alpha,\beta}\,\hat U_{\alpha\beta}\ket{\psi}.
\label{eq:select}
\end{equation}
Crucially, \textsc{Select} depends only on the fixed symbolic data, namely the left/right fragment dictionaries and their active support, and is therefore invariant under updates of the bridge \(\mathbf S_g(\vec\theta)\), whereas all coefficient dependence enters through \(\textsc{Prep}(\vec\theta)\). If the coefficients are complex, their signs or phases may be absorbed into the controlled unitaries in the standard LCU manner.

Defining the usual LCU sandwich operator
\begin{equation}
W(\vec\theta)
:=
\bigl(\textsc{Prep}^\dagger(\vec\theta)\otimes I\bigr)\,
\textsc{Select}\,
\bigl(\textsc{Prep}(\vec\theta)\otimes I\bigr),
\label{eq:W_def}
\end{equation}
the \(\ket{0}^{\otimes a}\)-block satisfies the block-encoding identity
\begin{equation}
(\bra{0}^{\otimes a}\otimes I)\,
W(\vec\theta)\,
(\ket{0}^{\otimes a}\otimes I)
=
\frac{\hat G(\vec\theta)}{\Lambda(\vec\theta)}.
\label{eq:block}
\end{equation}

If the goal is state preparation, one applies this block encoding to a reference state \(\ket{\phi_0}\) and then postselects the ancilla or uses amplitude amplification to obtain a state proportional to \(\hat G(\vec\theta)\ket{\phi_0}\). The corresponding postselection success probability is
\begin{equation}
p_{\rm succ}(\vec\theta)
=
\frac{\|\hat G(\vec\theta)\ket{\phi_0}\|^2}{\Lambda(\vec\theta)^2},
\end{equation}
so the usefulness of the construction depends not only on the reuse of the symbolic \textsc{Select} oracle but also on the normalization overhead encoded in \(\Lambda(\vec\theta)\).

Because \(\hat P^{\mathsf L}_\alpha\) and \(\hat P^{\mathsf R}_\beta\) act on disjoint subsystems, \textsc{Select} may also be implemented as commuting left/right components,
\begin{equation}
\textsc{Select}
=
\textsc{Select}_{\mathsf L}\textsc{Select}_{\mathsf R},
\end{equation}
which can be advantageous for modular compilation, although it is not required for correctness.

Equations~\eqref{eq:lcu_probs}--\eqref{eq:block} therefore define a direct map from the coefficient-decoupled MPO data
\begin{equation}
\bigl(
\mathcal G_{\mathsf L},
\mathcal G_{\mathsf R},
\mathcal E,
\mathbf S_g(\vec\theta)
\bigr)
\end{equation}
to an LCU-style \textsc{Prep}/\textsc{Select} interface. This supports two complementary workflows. First, offline-trained or classically optimized bridges \(\mathbf S_g(\vec\theta)\) may be loaded through a common oracle interface and used as state-preparation primitives inside subsequent quantum subroutines. Second, in hybrid classical–quantum loops, the symbolic \textsc{Select} circuit is compiled once from \(\mathcal G_{\mathsf L}\), \(\mathcal G_{\mathsf R}\), and \(\mathcal E\), while iterative coefficient updates are realized solely through updates of \(\textsc{Prep}(\vec\theta)\). This gives a clean separation between reusable symbolic compilation and dynamic coefficient loading.

\section{Conclusion and outlook}
\label{sec:conclusion}

We introduced a \emph{coefficient-decoupled} MPO framework that separates the symbolic operator structure of a many-body operator from a compact coefficient bridge across a chosen bipartition. This separation is useful both classically and quantum mechanically: on the classical side, it yields a reusable operator scaffold in which updates are confined to a small coefficient object, while on the quantum side, it induces a direct \textsc{Prep}/\textsc{Select} interface with static symbolic compilation and dynamic coefficient loading.

We further showed that coefficient decoupling provides a natural generator formalism \(\hat G(\vec\theta)\) for variational state preparation. Because the trial state depends linearly on the bridge coefficients, both ground-state search and target-state fitting reduce to effective environment contractions and low-dimensional linear-algebra problems at the cut. This makes the framework well suited to hybrid classical–quantum workflows, in which coefficients may be trained classically and then transferred to quantum hardware through the same oracle interface.

The coefficient bridge also provides a natural compression target, while the resulting LCU interface opens the door to integration with downstream block-encoding-based subroutines such as qubitization and quantum singular value transformation. Overall, coefficient-decoupled MPOs provide a compact and reusable interface between tensor-network representations and oracle-based quantum algorithms, combining controlled expressivity, efficient coefficient updates, and a direct pathway from classical compilation to quantum state preparation.

\section{Acknowledgement}
The author thanks Hendrik Weimer, Tobias J. Osborne, and Tanmoy Pandit for insightful discussions. The author acknowledges funding from the Ministry of Science and Culture of Lower Saxony through \textit{Quantum Valley Lower Saxony Q1} (QVLS-Q1).

\bibliographystyle{apsrev4-1}
\bibliography{references}

\appendix

\section{Symbolic compilation at a bipartition: left/right graphs and coefficient bridge}
\label{app:symbolic-graphs-bridge}

This appendix summarizes the symbolic compilation procedure underlying the coefficient-decoupled MPO construction. Given a Pauli-sum operator, the goal is to build reusable left and right symbolic graphs at a fixed cut and to assemble the sparse coefficient bridge coupling the two fragment dictionaries. In this way, the symbolic support is separated from the numerical weights.

\paragraph{Notation.}
We consider \(N\) qubits labeled \(0,1,\dots,N-1\). A Pauli string is written as
\[
P=\sigma_{p_0}\otimes\cdots\otimes\sigma_{p_{N-1}},
\qquad
p_i\in\{I,X,Y,Z\}.
\]
For indices \(a<b\), we write \(P[a{:}b]\) for the substring on sites \(a,a+1,\dots,b-1\). For a cut at \(M\), the left fragment is \(P[0{:}M]\) and the right fragment is \(P[M{:}N]\).

\subsection{Algorithm: left/right graphs and coefficient bridge}

\begin{algorithm}[H]
\caption{Left/right symbolic graphs and coefficient bridge}
\label{alg:lr_bridge}
\begin{algorithmic}[1]
    \STATE \textbf{input:} Pauli list \(H=\{(c_k,P^{(k)})\}_{k=1}^{K}\) on \(N\) sites.
    \STATE \textbf{output:} Symbolic graphs \(\mathcal{G}_L,\mathcal{G}_R\) and sparse bridge \(C\).
    \item[]
    \STATE \(M \gets \lfloor N/2 \rfloor\)
    \item[]
    \STATE \textbf{Phase 1: symbolic compilation}
    \STATE Initialize left layers \(\{\mathcal{V}^L_i\}_{i=0}^{M}\) with \(\mathcal{V}^L_0 \gets \{\epsilon\}\).
    \FOR{\(i=1\) \TO \(M\)}
        \STATE \(\mathcal{V}^L_i \gets \mathrm{Unique}\big(\{P^{(k)}[0{:}i] : k=1,\dots,K\}\big)\)
    \ENDFOR
    \STATE Initialize right layers \(\{\mathcal{V}^R_i\}_{i=0}^{N-M}\) with
    \[
    \mathcal{V}^R_0 \gets \mathrm{Unique}\big(\{P^{(k)}[M{:}N] : k=1,\dots,K\}\big).
    \]
    \FOR{\(i=1\) \TO \(N-M\)}
        \STATE \(\mathcal{V}^R_i \gets \mathrm{Unique}\big(\{P^{(k)}[M+i{:}N] : k=1,\dots,K\}\big)\)
    \ENDFOR
    \STATE \(\mathcal{G}_L \gets \mathrm{BuildGraph}\big(\{\mathcal{V}^L_i\}_{i=0}^{M}\big)\)
    \STATE \(\mathcal{G}_R \gets \mathrm{BuildGraph}\big(\{\mathcal{V}^R_i\}_{i=0}^{N-M}\big)\)
    \item[]
    \STATE \textbf{Phase 2: bridge assembly}
    \STATE Initialize sparse map \(C \gets 0\) on \(\mathcal{V}^L_M \times \mathcal{V}^R_0\).
    \FOR{\(k=1\) \TO \(K\)}
        \STATE \(u_k \gets P^{(k)}[0{:}M]\), \(\quad v_k \gets P^{(k)}[M{:}N]\)
        \STATE \(C[u_k,v_k] \gets C[u_k,v_k] + c_k\)
    \ENDFOR
    \item[]
    \STATE \textbf{return} \((\mathcal{G}_L, C, \mathcal{G}_R)\)
\end{algorithmic}
\end{algorithm}

Here \(\mathrm{BuildGraph}\) denotes the layered trie construction in which adjacent layers are connected by edges labeled by local symbols in \(\{I,X,Y,Z\}\). Repeated fragments are merged automatically, so the graphs record only distinct symbolic patterns, while the bridge stores the corresponding numerical couplings.

\subsection{Implementation remarks}
\label{app:bridge-remarks}

The bridge \(C\) may be stored either as a sparse dictionary keyed by fragment pairs \((u,v)\) or as a sparse matrix in a fixed index basis. In the latter case, introduce maps
\[
\mathrm{idx}_L:\mathcal{V}^L_M\to \{1,\dots,|\mathcal{V}^L_M|\},
\qquad
\mathrm{idx}_R:\mathcal{V}^R_0\to \{1,\dots,|\mathcal{V}^R_0|\},
\]
and represent the bridge as \(C_{\mathrm{idx}_L(u),\,\mathrm{idx}_R(v)}\). The graphs \(\mathcal G_L\) and \(\mathcal G_R\) can then be reused independently of subsequent coefficient updates.
\section{Compression of MPOs}
\label{app:mpo_compression}

After an MPO has been constructed, it is often compressed by bringing it into canonical form and discarding singular values below a chosen threshold. Without truncation, this produces an equivalent MPO; discarding only zero singular values gives an exact representation with reduced bond dimensions, while discarding small nonzero singular values yields a controlled approximation.

Consider an MPO
\begin{equation}
\hat O
=
\sum_{\substack{s_1,\ldots,s_N\\ s_1',\ldots,s_N'}}
W^{s_1 s_1'}
W^{s_2 s_2'}
\cdots
W^{s_N s_N'}
\ket{s_1\cdots s_N}\bra{s_1'\cdots s_N'}.
\end{equation}
Each local tensor \(W^{[j]}\) carries two virtual indices \(a_{j-1},a_j\) and two physical indices \(s_j,s_j'\).

Compression is performed by sweeping through the chain and reshaping each local tensor into a matrix, followed by QR or singular-value decompositions. In particular, one may first bring the MPO into right-canonical form by sweeping from right to left, and then into left-canonical or mixed-canonical form by sweeping from left to right.

The canonical conditions for an MPO tensor \(W^{[j]}\) are

\paragraph{Left-canonical.}
\begin{equation}
\sum_{a_{j-1},s_j,s_j'}
W^{s_j s_j' *}_{a_{j-1}a_j}
W^{s_j s_j'}_{a_{j-1}\tilde a_j}
= \delta_{a_j,\tilde a_j},
\end{equation}

\paragraph{Right-canonical.}
\begin{equation}
\sum_{a_j,s_j,s_j'}
W^{s_j s_j'}_{a_{j-1}a_j}
W^{s_j s_j' *}_{\tilde a_{j-1}a_j}
= \delta_{a_{j-1},\tilde a_{j-1}}.
\end{equation}

In a mixed-canonical representation, tensors to the left of a chosen bond satisfy the left-canonical condition, tensors to the right satisfy the right-canonical condition, and the singular values on the central bond encode the operator-Schmidt spectrum across that cut.

This connects directly to the coefficient-decoupled representation used in the main text. There, the bridge matrix \(\mathbf C=[C_{\alpha\beta}]\) already plays the role of a cut-local coefficient object. If the left and right fragment bases are orthonormal under the Hilbert–Schmidt inner product, then the singular value decomposition
\begin{equation}
\mathbf C = U \Sigma V^\dagger
\end{equation}
is the Schmidt decomposition of the operator coefficients across the chosen cut. Truncating to the leading \(r\) singular values gives the optimal rank-\(r\) approximation in Frobenius norm,
\begin{equation}
\mathbf C \approx \mathbf C_r = U_r \Sigma_r V_r^\dagger,
\end{equation}
and the retained rank \(r\) determines the effective bond dimension across the cut.

Writing
\begin{equation}
\mathbf C_r = U_r \Sigma_r^{1/2}\,\Sigma_r^{1/2} V_r^\dagger,
\end{equation}
one may absorb \(U_r \Sigma_r^{1/2}\) into the left symbolic side and \(\Sigma_r^{1/2} V_r^\dagger\) into the right symbolic side. In this way, standard MPO compression and bridge compression are the same idea viewed at different levels: the former is implemented by canonicalization and truncation along the chain, while the latter concentrates the compression explicitly at the chosen cut.
\subsection{Symbolic matrix product operator construction via cut matrices and pivoted QR}
\label{app:symbolic-mpo-qr}

This appendix derives a standard symbolic matrix product operator construction routine based on cut coefficient matrices and rank-revealing (column-pivoted) QR factorizations. It clarifies how Pauli-sum coefficients are absorbed into local matrix product operator tensors during a left-to-right sweep and contrasts this with the coefficient-bridge representation in the main text, where symbolic support is fixed and coefficients are isolated.

\subsubsection{Setup: Pauli-sum Hamiltonian}
We consider a Pauli sum on \(N\) qubits
\begin{equation}
H=\sum_{\alpha=1}^{K} c_\alpha\,P_\alpha,
\qquad
P_\alpha=\bigotimes_{i=0}^{N-1}\sigma_{p_{\alpha,i}},
\qquad
p_{\alpha,i}\in\{I,X,Y,Z\}.
\label{eq:pauli-sum-app}
\end{equation}
At sweep step \(i\) (cut after site \(i\)), each term splits as
\begin{equation}
P_\alpha
=
\Big(\bigotimes_{m=0}^{i}\sigma_{p_{\alpha,m}}\Big)
\otimes
\Big(\bigotimes_{m=i+1}^{N-1}\sigma_{p_{\alpha,m}}\Big)
\equiv
L^{(i)}_\alpha\otimes R^{(i)}_\alpha.
\label{eq:split-LR}
\end{equation}

\subsubsection{Cut coefficient matrix \(\Gamma^{(i)}\)}
After deduplicating the left patterns \(\{L^{(i)}_j\}_{j=1}^{J_i}\) and right patterns \(\{R^{(i)}_k\}_{k=1}^{K_i}\), the Hamiltonian can be written as
\begin{equation}
H=\sum_{j=1}^{J_i}\sum_{k=1}^{K_i}\Gamma^{(i)}_{jk}\,L^{(i)}_j\otimes R^{(i)}_k,
\label{eq:cut-expansion-app}
\end{equation}
where the cut coefficient matrix \(\Gamma^{(i)}\in\mathbb{C}^{J_i\times K_i}\) is
\begin{equation}
\Gamma^{(i)}_{jk}
=
\sum_{\alpha:\,L^{(i)}_\alpha=L^{(i)}_j,\;R^{(i)}_\alpha=R^{(i)}_k}
c_\alpha.
\label{eq:gamma-def-app}
\end{equation}

\subsubsection{Pivoted QR and the local tensor \(W^{[i]}\)}
A rank-revealing step is obtained via column-pivoted QR,
\begin{equation}
\Gamma^{(i)} P^{(i)} = Q^{(i)} R^{(i)},
\qquad\Rightarrow\qquad
\Gamma^{(i)} = Q^{(i)}\Big(R^{(i)}(P^{(i)})^\top\Big),
\label{eq:pivoted-qr-app}
\end{equation}
where \(P^{(i)}\) is a permutation matrix. In practice, one may choose a numerical rank \(r_i\) by a tolerance on the diagonal of \(R^{(i)}\) (or an equivalent criterion) and truncate to obtain
\begin{equation}
\Gamma^{(i)} \approx Q^{(i)}_{(:,1:r_i)}\,\widetilde{\Gamma}^{(i)},
\qquad
\widetilde{\Gamma}^{(i)}:=R^{(i)}_{(1:r_i,:)}(P^{(i)})^\top,
\label{eq:qr-trunc-app}
\end{equation}
thereby introducing an outgoing bond index \(\ell\in\{1,\dots,r_i\}\) with bond dimension \(\chi_{i+1}=r_i\).

To connect \(Q^{(i)}\) to the matrix product operator tensor, write each unique left pattern at step \(i\) as
\begin{equation}
L^{(i)}_j = B_{a(j)}\otimes \sigma_{p(j)},
\label{eq:left-pattern-Bp}
\end{equation}
where \(a(j)\in\{1,\dots,\chi_i\}\) labels the incoming bond-basis operator \(B_a\) and \(p(j)\in\{I,X,Y,Z\}\) is the local Pauli symbol at site \(i\). Grouping by the same incoming index \(a\) yields
\begin{equation}
\widetilde{L}^{(i)}_\ell
=
\sum_{j=1}^{J_i}Q^{(i)}_{j\ell}\,L^{(i)}_j
=
\sum_{a=1}^{\chi_i}
B_a\otimes
\left(
\sum_{j:\,a(j)=a}Q^{(i)}_{j\ell}\,\sigma_{p(j)}
\right),
\end{equation}
which identifies the operator-valued tensor entries as
\begin{equation}
W^{[i]}_{a\ell}
=
\sum_{j:\,a(j)=a}Q^{(i)}_{j\ell}\,\sigma_{p(j)}
\;=\;
\sum_{p\in\{I,X,Y,Z\}} w^{[i]}_{a\ell}(p)\,\sigma_p,
\label{eq:W-from-Q-app}
\end{equation}
with coefficients
\begin{equation}
w^{[i]}_{a\ell}(p)=\sum_{j:\,a(j)=a,\;p(j)=p} Q^{(i)}_{j\ell}.
\label{eq:w-coeffs-app}
\end{equation}
The residual matrix \(\widetilde{\Gamma}^{(i)}\) is then carried forward as a new suffix table on sites \(i+1,\dots,N-1\). The cost of this method is essentially the same as first building the symbolic MPO and then compressing it, since both approaches manipulate objects of comparable size. The difference is that here the compression is performed on the fly during construction, whereas in the standard workflow, it is applied afterward.

\subsubsection{Example (subset of $\mathrm{H}_2$ on 4 qubits)}
\label{app:h2-worked-example}

We illustrate the cut-matrix construction on \(N=4\) using a subset of \(9\) Pauli terms from the \(\mathrm{H}_2\) Hamiltonian at \(R=0.7414\)\,\AA\ (Table~\ref{tab:H2-subset-terms}).

\paragraph{First cut (\(i=0\)): explicit \(\Gamma^{(0)}\)}
At \(i=0\), the unique left patterns are \(\{I_0,Y_0,Z_0\}\) (no \(X_0\) appears), and the unique right suffixes on qubits \(1\)--\(3\) are
\[
\{\,III,\ IZI,\ ZII,\ ZZI,\ XXY,\ XYY,\ IZZ,\ IIZ\,\}.
\]
Using row order \([I,Y,Z]\) and column order \([III,\,IZI,\,ZII,\,ZZI,\,XXY,\,XYY,\,IZZ,\,IIZ]\), one obtains
\begin{widetext}

\begin{table}[h]
    \centering
    \caption{Subset of \(\mathrm{H}_2\) Pauli terms used for the worked example. Coefficients are rounded for readability.}
    \label{tab:H2-subset-terms}
    \begin{tabular}{lc @{\hspace{1.5em}} lc}
        \toprule
        \textbf{Term} & \textbf{Coefficient} & \textbf{Term} & \textbf{Coefficient} \\
        \midrule
        \(IIII\) & \(-0.098864\) & \(XXYY\) & \(-0.045322\) \\
        \(ZIII\) & \(+0.171198\) & \(ZIZI\) & \(+0.120545\) \\
        \(IIZI\) & \(-0.222786\) & \(IZZI\) & \(+0.165867\) \\
        \(ZZII\) & \(+0.168622\) & \(IIZZ\) & \(+0.174348\) \\
        \(YXXY\) & \(+0.045322\) & & \\
        \bottomrule
    \end{tabular}
\end{table}

\begin{equation}
\Gamma^{(0)}=
\begin{pmatrix}
-0.098864 & -0.222786 & 0 & +0.165867 & 0 & 0 & +0.174348 & 0 \\
0 & 0 & 0 & 0 & +0.045322 & 0 & 0 & 0 \\
+0.171198 & 0 & +0.168622 & 0 & 0 & +0.120545 & 0 & 0
\end{pmatrix}.
\label{eq:Gamma0-H2-subset}
\end{equation}
\end{widetext}

A pivoted QR \(\Gamma^{(0)}P^{(0)}=Q^{(0)}R^{(0)}\) sets \(\chi_1=r_0\le 3\) and defines the first MPO tensor \(W^{[0]}\), whose local operator entries are expanded in the basis \(\{I,X,Y,Z\}\) according to Eqs.~\eqref{eq:W-from-Q-app}--\eqref{eq:w-coeffs-app}. Since \(\chi_0=1\), each operator-valued entry can be written explicitly as
\begin{equation}
W^{[0]}_{1\ell}
=
q_{I,\ell}\,I_0 + q_{Y,\ell}\,Y_0 + q_{Z,\ell}\,Z_0,
\end{equation}
where \((q_{I,\ell},q_{Y,\ell},q_{Z,\ell})^\top\) is column \(\ell\) of \(Q^{(0)}\) in the chosen row ordering.

\paragraph{Second cut (\(i=1\)): formation of \(\Gamma^{(1)}\)}
After absorbing site \(0\), the residual suffix table on sites \(1\)--\(3\) is updated to
\(\widetilde{\Gamma}^{(0)}=R^{(0)}(P^{(0)})^\top\) (or its rank-\(r_0\) truncation). At \(i=1\), the unique left patterns are pairs \((a,p)\) with incoming bond label \(a\in\{1,\dots,\chi_1\}\) and local Pauli symbol \(p\in\{I,X,Y,Z\}\) on site \(1\), so \(L^{(1)}_j=B_{a(j)}\otimes\sigma_{p(j)}\). Deduplicating these patterns and the remaining suffixes on sites \(2\)--\(3\) yields \(\Gamma^{(1)}\in\mathbb{C}^{J_1\times K_1}\) via Eq.~\eqref{eq:gamma-def-app}. The pivoted QR
\(\Gamma^{(1)}P^{(1)}=Q^{(1)}R^{(1)}\) then defines the next MPO tensor \(W^{[1]}\in\mathbb{C}^{\chi_1\times\chi_2}\) through the grouping rule \eqref{eq:W-from-Q-app}.

\paragraph{Remark (relation to coefficient-decoupled representation).}
In the standard sweep construction above, numerical coefficients are redistributed between local tensors and the carried suffix table through the QR steps \eqref{eq:pivoted-qr-app}. In the coefficient-decoupled representation in the main text, the left/right symbolic graphs are fixed, and the tunable numerical degrees of freedom are isolated in an explicit bridge object, which is convenient for coefficient updates and for interfacing with \textsc{Prep}/\textsc{Select} LCU oracles.

\section{Electronic-structure preliminaries and fermion-to-qubit mapping}
\label{app:chem_prelim}

We summarize the standard electronic-structure notation used in Sec.~\ref{sec:app_chem}. In an orthonormal spin-orbital
basis $\{\varphi_p(x)\}$, the second-quantized electronic Hamiltonian is
\begin{equation}
    \hat{H} = \sum_{pq} h_{pq}\, \hat a_p^\dagger \hat a_q
    + \frac{1}{2}\sum_{pqrs} g_{pqrs}\, \hat a_p^\dagger \hat a_q^\dagger \hat a_r \hat a_s ,
    \label{eq:fermionic_H_app}
\end{equation}
with canonical fermionic anticommutation relations. After computing integrals and a mean-field reference state
classically, we map fermionic operators to qubits to obtain a Pauli-sum Hamiltonian of the form in Eq.~\ref{eq:H_pauli} of the main text.

\section{Jordan--Wigner transformation}
\label{app:jw}

For $N$ spin orbitals mapped to $N$ qubits, define $\sigma_p^{\pm}=(X_p \mp iY_p)/2$. The Jordan-Wigner mapping is
\begin{equation}
\hat a_p
= \left(\prod_{m=0}^{p-1} Z_m\right)\sigma_p^-,
\qquad
\hat a_p^\dagger
= \left(\prod_{m=0}^{p-1} Z_m\right)\sigma_p^+ .
\end{equation}
The number operator is $\hat n_p=\hat a_p^\dagger \hat a_p = (I-Z_p)/2$. 
\section{Solving the effective generalized eigenproblem with LOBPCG}
\label{app:lobpcg}

To optimize coefficients in the effective small problem, we solve
\begin{equation}
H_{\rm eff} X = N_{\rm eff} X \Lambda,
\label{eq:app_gen_eig}
\end{equation}
where $H_{\rm eff}$ is Hermitian and $N_{\rm eff}$ is Hermitian positive definite (or regularized as
$N_{\rm eff}\leftarrow N_{\rm eff}+\lambda I$). We employ a matrix-free block iterative eigensolver such as LOBPCG,
since $H_{\rm eff}$ and $N_{\rm eff}$ are assembled from tensor-network contractions. A small block size
$b\in[2,8]$ is typically sufficient for the lowest Ritz vector in the coefficient-only update loop.

\subsection{Fidelity-based bridge optimization}
\label{sec:fidelity_bridge_optimization}

The same coefficient-bridge framework can also be used for target-state fitting. Given a target state \(\ket{\psi_{\rm tgt}}\), one may minimize
\begin{equation}
\min_x\ \bigl\|\ket{\psi(x)}-\ket{\psi_{\rm tgt}}\bigr\|^2,
\end{equation}
which yields the linear system
\begin{equation}
(N_{\rm eff}+\eta I)x=b,
\qquad
b_j:=\braket{\varphi_j|\psi_{\rm tgt}}.
\end{equation}
Here \(N_{\rm eff}\) is the Gram matrix from Eq.~\eqref{eq:heff_neff_defs} and \(\eta\ge 0\) is an optional regularization parameter. We do not explore this variant further, but it provides a straightforward extension of the bridge-based optimization framework.

\section{Perfect sampling from an MPS}
\label{app:perfect_sampling}

This appendix records the conditional perfect-sampling algorithm used to generate Pauli strings from a reference MPS. The algorithm samples from the target distribution \(\Pi(P)\) defined in Sec.~\ref{sec:pauli_conditional_sampling_mps}.

\begin{algorithm}[H]
\caption{Conditional perfect sampling of Pauli strings from an MPS}
\label{alg:pauli_sampling_mps}
\begin{algorithmic}[1]
    \STATE \textbf{input:} Right-canonical MPS tensors $\{A^{s_j}\}_{j=1}^{N}$ and number of samples $\mathcal{N}_{samp}$.
    \STATE \textbf{output:} Samples $\{P^{(\mu)}\}_{\mu=1}^{\mathcal{N}_{samp}}$, where $P^{(\mu)} \in \{I,X,Y,Z\}^{\otimes N}$.
    \item[]
    \FOR{$\mu=1$ \TO $\mathcal{N}_{samp}$}
        \STATE Set $E=\mathbb{I}$ and initialize $P^{(\mu)}$ as an empty string.
        \FOR{$j=1$ \TO $N$}
            \STATE Compute $\widetilde{\pi}_j(\alpha)$ for $\alpha \in \{I,X,Y,Z\}$.
            \STATE Set $\pi_j(\alpha)=\widetilde{\pi}_j(\alpha)\big/\sum_{\beta \in \{I,X,Y,Z\}} \widetilde{\pi}_j(\beta)$.
            \STATE Sample $\sigma_j \sim \pi_j(\cdot)$ and append $\sigma_j$ to $P^{(\mu)}$.
            \STATE Update $E$ using $A^{s_j}$, $A^{s_j\dagger}$, and $\sigma_j$.
        \ENDFOR
    \ENDFOR
    \item[]
    \STATE \textbf{return:} $\{P^{(\mu)}\}_{\mu=1}^{\mathcal{N}_{samp}}$.
\end{algorithmic}
\end{algorithm}

\end{document}